# Light Induced Degradation of $CH_3NH_3PbI_3$ Thin Film

*Youzhen Li,[a] Xuemei Xu,[a] Congcong Wang,[b] Ben Ecker,[b] Junliang Yang,[a] Jinsong Huang,[c] and Yongli Gao[b,]\**

[a]School of Physics and Electronics, Central South University, Changsha, Hunan, 410083, P.R. China

[b]Department of Physics and Astronomy, University of Rochester, Rochester, NY 14627，USA

[c]Department of Mechanical and Materials Engineering, Nebraska Center for Materials and Nanoscience, University of Nebraska-Lincoln, Lincoln, Nebraska, 68588, USA

**Corresponding author:Yongli Gao    Email:ygao@pas.rochester.edu**

# Light Induced Degradation of $CH_3NH_3PbI_3$ Thin Film

**ABSTRACT**

The stability of $CH_3NH_3PbI_3$ was investigated by observing the damage on a co-evaporated film irradiated by a blue laser in ultrahigh vacuum. X-ray photoelectron spectroscopy (XPS) and scanning electron microscopy (SEM) were employed to investigate the irradiation effects on the surface. The core levels of perovskite were observed to shift toward higher binding energy (BE) during the irradiation, suggesting that surface became more n-type. Metallic Pb appeared after 120 minutes of irradiation, indicating that the film has started to decompose. The decomposition saturated after about 480 minutes of irradiation when the ratio of metallic Pb to total Pb was about 33%. Furthermore, the film was no longer continuous after irradiation because of the stress change after irradiation, as the substrate Au and oxygen were detected by XPS. SEM images also show a roughened surface after irradiation. The results strongly indicate that the perovskite is sensitive to the laser irradiation, and the decomposition is a self-limiting process.

# 1. INTRODUCTION

Organic photovoltaics,[1] dye-sensitized solar cells,[2] and colloidal nanocrystal solar cells[3] have widely been investigated in recent years as an important source of renewable energy. The most critical factors for the utility of solar cells are their cost of production, power conversion efficiency (PCE), and the sustainability of the source materials. Recently organic-inorganic halide perovskites have attracted considerable attention as a light-harvesting material for new solar cells,[4-9] offering both low cost and reasonably high PCE compared with that of silicon, dye-sensitized, and other standard solar cells. Perovskite was first used in a dye-sensitized solar cell by Miyasaka and co-workers in 2009 with a PCE of 3.8%.[10] Kim and co-workers later reported a solid-state perovskite solar cell with a PCE of 9.7% in 2012.[5] Soon after that, perovskite solar cells with a planar heterojunction structure were also reported with a PCE of 5%,[11] and have since continued to rapidly increase to as high as certified value of 22.1%.[12] Perovskite solar cell's PCEs are presently quite close to that of the standard crystalline silicon solar cells.

Perovskite films can be fabricated by either solution based spin coating or thermal evaporation. The spin coating method is a simpler process and of lower production cost than that of the latter, and is therefore popularly used for device fabrication. Surface and interface investigations of spin-cast films have produced information important for understanding the performance of device fabricated with spin casting.[13-17] However, the surface composition of spin coated films usually deviates significantly from the stoichiometry ratio.[18,19] The thermal evaporation method can produce smoother films with the correct surface stoichiometry and crystal structure,[20-22] which is essential for surface sensitive analytical investigations on the fundamental materials properties of the perovs

Although the PCE of perovskite solar cell has already reached over 20% in laboratory,[12] the stability of the light-harvesting film ($CH_3NH_3PbX_3$, X = I, Cl, Br) remains a critical issue for the eventual success of commercial applications. Early work attributes illumination induced degradation of perovskite solar cells to the $TiO_2$ layer in the device.[23] Murugadoss *et al.* investigated light illumination stability of thin perovskite films on different substrates and found a strong substrate dependence.[24] Bryant *et al.* demonstrated light illumination played a key role in perovskite degradation when exposed to gases and ambient.[25] Applying simultaneous luminescence and electron microscopy, Yuan *et al.* found that light-induced degradation results in material decomposition but little photoluminescent (PL) spectral shifts.[26] On the other hand, Merdasa *et al.* observed rapid degradation and continuous PL spectral blue shift of perovskite nanocrystals using super-resolution luminescence microspectroscopy.[27] We have previously shown that $CH_3NH_3PbI_3$ is sensitive to moisture, decomposing rapidly after reaching a threshold of $2 \times 10^{10}$ Langmuir of water exposure.[22]

In this work, we present our investigation of the effects of laser irradiation on $CH_3NH_3PbI_3$ films using XPS and SEM. The films were prepared by thermal co-evaporation, with proper stoichiometry, uniformity, and crystal structure. After laser irradiation for only 120 minutes by blue light of wavelength 408 nm at intensity ~7 times of AM 1.5, a new component of Pb started to appear, marking the initiation of perovskite decomposition. The decomposition saturated after about 480 minutes of irradiation at the same intensity, with the perovskite Pb reduction by about 33%. The uniformity of the film was also destroyed by the irradiation, as the substrate Au and oxygen were detected by XPS and roughening of the surface was shown by SEM. The results provide intriguing insight of the complicated degradation process by irradiation.

## 2. EXPERIMENTAL SECTION

***Film fabrication:*** The film evaporation and XPS measurements were performed in a modified Surface Science Laboratories' SSX-100. A 10 nanometer thick $CH_3NH_3PbI_3$ film was thermally evaporated onto an Au coated silicon substrate in a sample preparation chamber attached to the analytical chamber. The details of the instrument and the evaporation procedure are described in Ref. 28.

***Irradiation and XPS measurements:*** The sample was then transferred into the analytical chamber of base pressure $1\times10^{-10}$ torr. The irradiation light used for exposure came from a semiconductor laser with a wavelength of 408 nm. The laser's output power was 20 mW and the irradiation intensity on the sample was about $6.8\times10^{-3}$ $W/mm^2$, about 7 times of the commonly used AM1.5 irradiation ($1000W/m^2$). The laser irradiation covers an oval shaped area approximately 1.4 x 2.6 mm, and the probing x-ray spot is about 0.8 x 1.4 mm in size. After each timed irradiation, we immediately performed XPS measurements on both the laser irradiated and another non-irradiated spots on the surface. Precise control of the sample positions were guided by an optical microscope attached on the chamber. Other windows on the chamber were all shaded to avoid the light influence from the outside.

***Data process and SEM images:*** The core level peaks were fitted with commercial software Origin 6.0, and the ratio of Lorentzian and Gaussian was not fixed during the peak fitting. The atomic ratio of the film was calculated by the areas of fitted curves divide by the sensitivity factor. Shirley-type corrections were performed to remove the secondary electron background. At the end of the experiment with a total irradiation time of 2210 min., the sample was taken out of the chamber and was investigated immediately with Zeiss Auriga SEM located at the University of Roch

UR-nano center.

## 3. RESULTS AND DISCUSSION

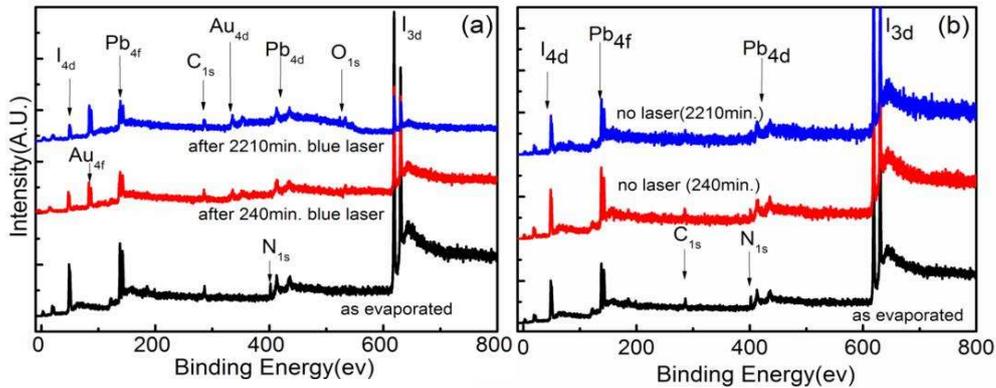

**Figure 1.** Survey XPS spectrum of $CH_3NH_3PbI_3$ film:（a）laser irradiated position and (b)non-irradiated position. Some substrate Au was exposed by the irradiation, not seen at the non-irradiate position.

Figure 1(a) shows the survey scan of the film before and after laser irradiation. For comparison, the scan of non-irradiated position is also shown in Figure 1(b). There were no other obvious elements in the as evaporated survey scan except for C, N, I, and Pb, indicating that the $CH_3NH_3PbI_3$ film was uniform and free of contamination.  The atomic ratios of the as-evaporated film are: N: I: Pb = 1.10:3.28:1.00 and 1.09:3.26:1.00 at the laser illuminated and non-illuminated spot, respectively. Carbon was not shown here because the contamination from the substrate is difficult to eliminate. The ratio of N: I: Pb can demonstrate a reasonably good stoichiometry.  The atomic ratio also shows the surface uniformity.  The crystallinity of the film thus prepared has been demonstrated previously.[29] As shown in Figure 1(a), after only 240 minutes of irradiation, the Au 4f and O 1s peaks were clearly visible, indicating that the film was no longer uniform.  The presence of Au and O suggests that the photo-emitted electrons were coming from the underlying gold substrate, which had water or oxygen adsortbates on it, because it was not sputter cleaned prior to the perovskite evaporation.  At the end of the experiment when the sample was irradiated for 2210 minut intensity of the Au 4f and O 1s peaks further increased. In contrast, the XPS spectra from the non-irradiated position show no gold or oxygen, as can be seen in Figure 1(b). The XPS survey scan results strongly suggest that the $CH_3NH_3PbI_3$ film is sensitive to laser irradiation, and the high intensity (~7 times of AM 1.5) of the laser may accelerate the degradation progress. In addition, we found that after only about 120 minutes of laser irradiation, the color of the exposed spot became more visibly lighter than that of the non-irradiated region.

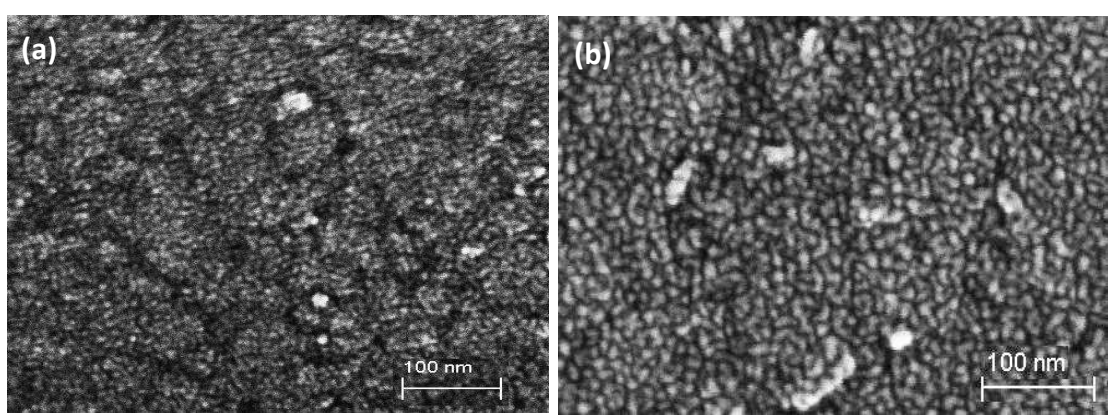

**Figure 2.** SEM images of the co-evaporated film at (a) after 2120 min. laser irradiation, and (b) non-irradiated position. The irradiated region is much rougher than the non-irradiated one.

Figure 2 shows the top-view SEM of the co-evaporated $CH_3NH_3PbI_3$ film. Figure 2(a) shows that after 2210 min. of laser irradiation, the film was roughened. There are small cracks indicating that the film was not uniform. These cracks by irradiation should be responsible for the exposure of the substrate Au observed in XPS. The SEM image of the non-irradiated position shown in Figure 2(b) indicates the film is uniform and with a polycrystalline structure. There are some bright linear structures with a length of less than one hundred nanometer, which may come from excessive $CH_3NH_3I$ particles. The obviously change of the surface morphology before and after irradiation also indicates the degradation by irradiation. The evaporated $CH_3NH_3PbI_3$ films have been demonstrated to be more uniform and efficient than solution-processed ones.[9]

In Figure 3(a) and (b) we present the valence band of the $CH_3NH_3PbI_3$ film durin experiment. Figure 3(a) is from the laser irradiated position, while Figure 3(b) is from the non-irradiated one. We found that the valence band maximum (VBM) increased by about 0.4 eV, from 0.85 eV initially to 1.25 eV after the total 2210 minutes of irradiation. The shift seemed already saturated after about 480 minutes of irradiation. The band gap of $CH_3NH_3PbI_3$ is known to be around 1.55 eV,[17] so the "as prepared film" was a slightly n-type one. The shift towards a high binding energy (BE) indicated the film became more n-type during the laser irradiation. Meanwhile at the non-irradiated position no obvious changes can be seen in the VBM region. Another feature to notice is the change of the spectra by the laser irradiation. After only 120 minutes of laser irradiation, a peak at about 6.5 eV begins to appear. This indicates the existence of oxygen on the surface. As the film is in the ultrahigh vacuum chamber and is not sensitive to oxygen exposure,[22] the oxygen could only come from contaminants on the underlying substrate since the substrate surface was not cleaned by Ar sputtering prior to evaporation of the perovskite film. This means that the film was already not continuous as photo-emitted electrons from the underlying gold substrate could be seen.

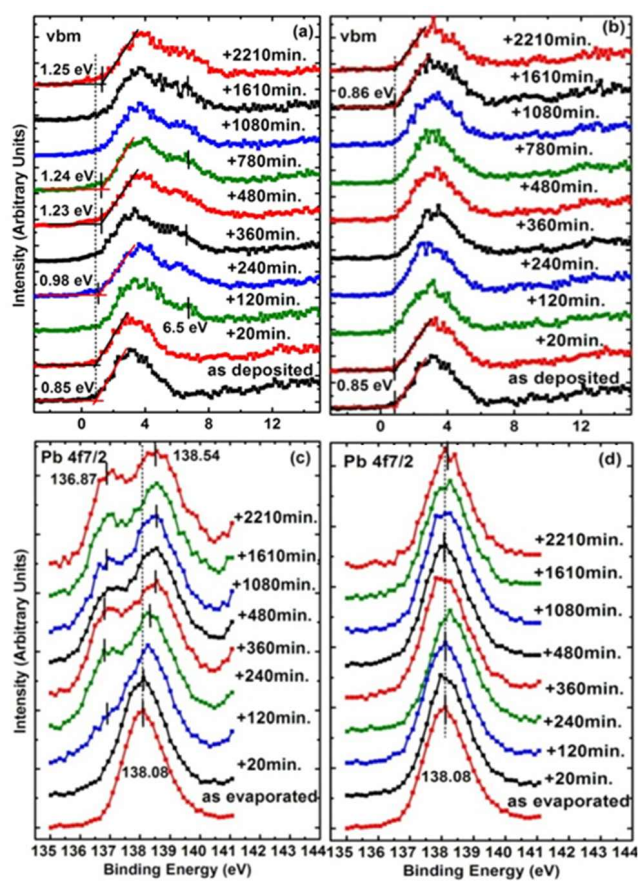

**Figure 3.** XPS spectra of valence band and Pb $4f_{7/2}$ of the $CH_3NH_3PbI_3$ film at (a) (c) laser irradiated position and (b) (d) non-irradiated position. The irradiation induced changes are clearly visible in (a) and (c) after the irradiation.

Shown in Figure 3(c) and (d) are the Pb $4f_{7/2}$ spectra from laser irradiated and non-irradiated position, respectively. The BE of as-evaporated Pb $4f_{7/2}$ is about 138.08 eV, consistent to the previous report.[28] As the laser irradiation progresses, changes can be observed in the core level. After about only 120 min. laser irradiation, a new peak at about 136.87 eV in Pb $4f_{7/2}$ began to appear (Figure 3(c)), and grew in intensity as the irradiation continued. The new Pb $4f_{7/2}$ peak is that of metallic Pb, and it seems synchronized to the 6.5 eV peak in the VBM region. In contrast, the Pb $4f_{7/2}$ spectra of no laser irradiated position shown in Figure 3(d) present no discernable spectral change.

$CH_3NH_3PbI_3$ films are usually known to decompose into $PbI_2$ and other components.[21,29] $PbI_2$ is photosensitive and can usually decompose into PbO and $I_2$ in the presence of moisture and light.[30] In our case, the film was in ultra-high vacuum(UHV), where there was no $H_2O$ or oxygen. It is apparent that the $PbI_2$ released by laser irradiation further decomposes into metallic lead and iodine by irradiation alone, leading to the new peak in Pb $4f_{7/2}$ at 136.87 eV as shown in Figure 3(c). Iodine released in this photochemical reaction will sublimate to vacuum, leaving the metallic Pb with the remaining perovskite. Previous reports have also found metallic lead in as fabricated $CH_3NH_3PbI_3$ films (especially for spin cast ones), and it was considered to be related with the excessive precursor $PbI_2$ and the removal of $I_2$.[18] The binding energy of metallic Pb is reporte 137.0 eV, [31-33] which is very close to that of the new component at 136.87 eV after irradiation. The metallic lead will act as quenching centers of excitons, evidenced by the detrimental effect on the photoluminescence quantum efficiency.[33] As a result, this degradation process will have a serious effect on the device PCE. On the other hand, the degradation process may be kinematically limited in the bulk of perovskite or in a well sealed device when the volatile components in the reaction are not allowed to escape as easily as in a vacuum.

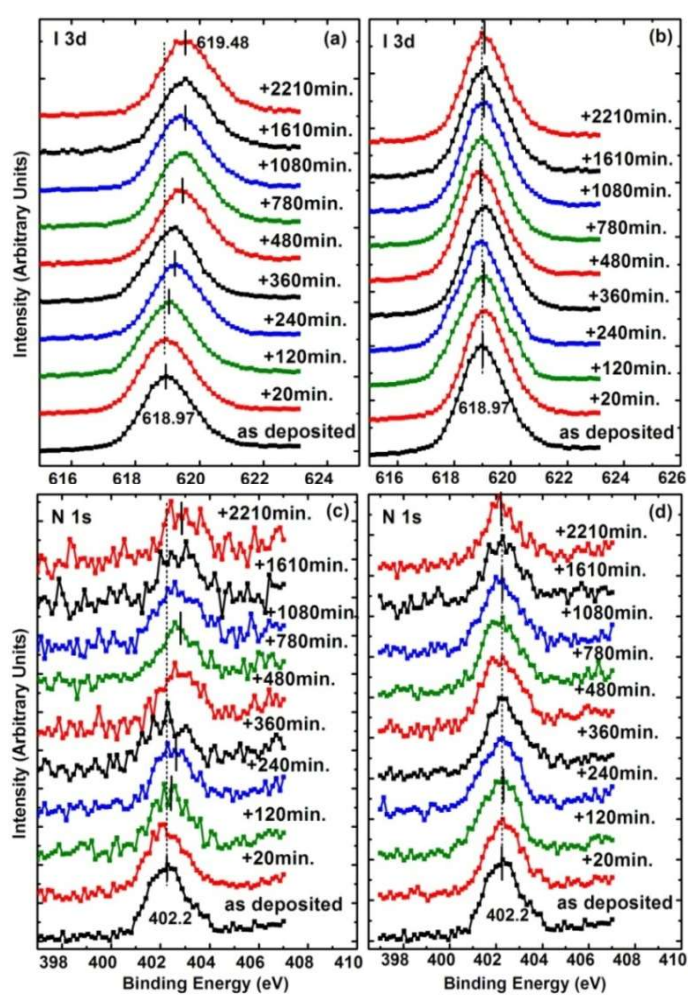

**Figure 4** XPS spectra of I 3d and N 1s of the $CH_3NH_3PbI_3$ film at
(a) (c) laser irradiated position and (b) (d) non-irradiated position.

Beside the perovskite decomposition, we also found a ~0.46 eV shift of perovskite Pb $4f_{7/2}$ from 138.08 to 138.54 eV in the irradiated position (see Figure 3(c)). This shift of the perc Pb $4f_{7/2}$ is consistent with that of the VBM, supporting the notion that the perovskite has b more n-doped. The n-doping may be due to the face that the film has become Pb rich, similar to the case when perovskite is annealed at high temperature.[18]

Further support of the n-doping of the perovskite can be obtained by observing the shifts of other core levels. Should there be n-doping, the movement of the Fermi level within the band gap toward the conduction band minimum will result in a rigid shift in all the energy levels in the semiconductor. The evolution of I $3d_{5/2}$ and N 1s at laser irradiated (Figure 4(a) and (c)) and non-irradiated (Figure 4(b) and (d)) position are both shown in Figure 4 to have a comparison. Dotted lines and short bars are marked to show the core level shift. For the laser irradiated position, the I $3d_{5/2}$ had a ~0.51 eV shift (Figure 4(a)) towards higher binding energy from 618.97 eV to 619.48 eV after the irradiation. The shift seems already saturated after 480 min. irradiation. Similar shift can also be seen for N 1s in Figure 4(c). The shift is about the same as that of Pb $4f_{7/2}$ considering the instrument uncertainty. We can conclude that the shift is caused by the shift of the Fermi level of the film, consistent with the VBM shift shown in Figure 3(a). We also found a serious intensity decrease of N 1s after irradiation (Figure 4(c), indicating the decomposition of the film and the escape of N. For the non-irradiated position, such a shift is again not observed (Figure 4(b) and (d)).

Based on the above discussions, we deduce that the decomposition by irradiation can be divided into two steps. Firstly, $CH_3NH_3PbI_3$ decomposed into $PbI_2$ and other volatile components, and then $PbI_2$ further decomposed into metallic Pb and I. The degradation procedure can be proposed as the following:

$$CH_3NH_3PbI_3 \xrightarrow{light} (-CH_2-) + NH_3 + HI + PbI_2 \quad (1)$$

$$PbI_2 \xrightarrow{light} Pb + I_2 \quad (2)$$

NH$_3$ and HI should fly away from the surface, and so should the iodine by sublimation, leaving unreacted perovskite, metallic Pb, and carbon hydride complex on the surface. Without a means of stabilizing the perovskite thin film, the degradation will severely limit the ability of this material to be used in the consumer market as a light absorbing layer in a solar cell.

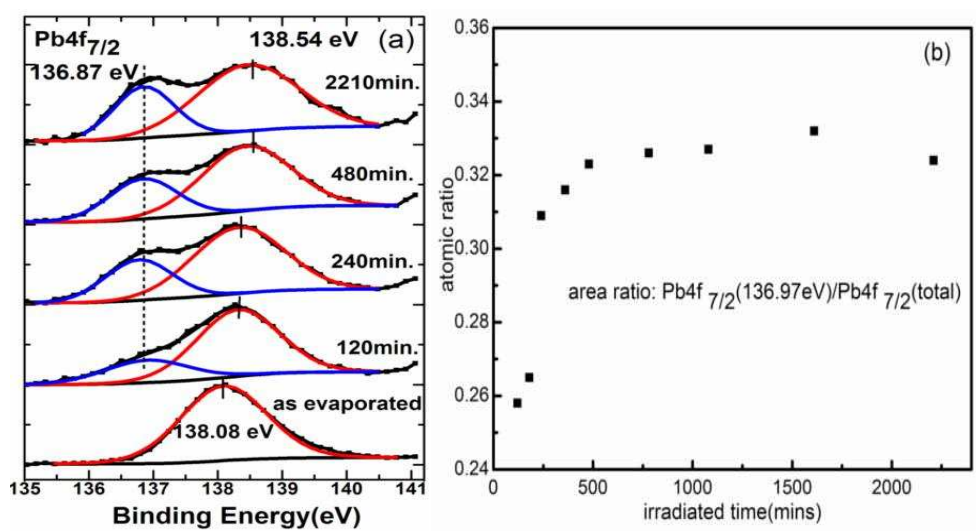

**Figure 5.** (a) Pb 4f$_{7/2}$ decomposition and (b) metallic Pb fraction during laser irradiation

The detailed curve fitting of Pb 4f$_{7/2}$ is shown in Figure 5(a). Dotted line and short bars clearly shows the appearance of metallic Pb at 136.87 eV and the 0.46 eV shift of perovskite Pb 4f$_{7/2}$ toward higher binding energy. It's clear that the shift of the perovskite Pb 4f$_{7/2}$ is already saturated after about 480 min. irradiation, in contrast to the stationary metallic one. Shown in Figure 5(b) is the ratio between metallic Pb and total Pb during the laser irradiation, calculated from the areas under the fitting curves in Figure 5(a). It increased from 26% after 120 min. to 33% after 480 min. irradiation, and then the ratio remained almost constant. This indicated that the degradation saturated after laser irradiation of about 480 minutes.

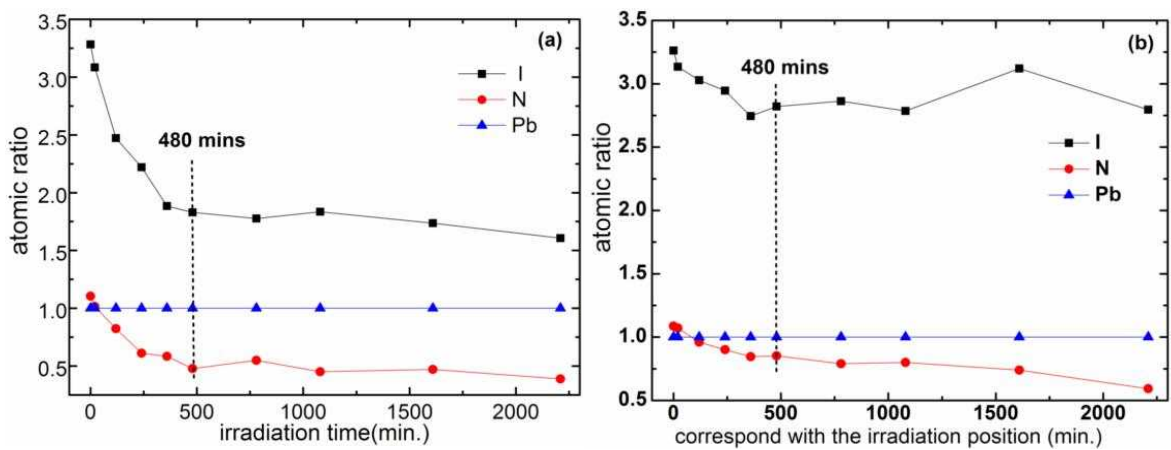

**Figure 6.** Atomic ratio change of (a) laser irradiated position and (b) non- irradiated position.

Figure 6 (a) and (b) show the atomic ratio change of the laser irradiated position and the non-irradiated position, respectively. For the laser exposed position, the ratio of N and I decreased quickly, from N: I: Pb=1.10:3.28:1.00 to 0.48:1.82:1.00 after 480 min. irradiation, at which point the ratios became relatively constant. For the non-exposed position, the ratio of N and I also decreased slightly, from N: I: Pb=1.09:3.26:1.00 to 0.85:2.82:1.00 at the time corresponding to 480 min. irradiation. The small reduction of the N:Pb ratio at the non-irradiated position is likely be caused by the probing x-ray. Reduction by extended exposure to vacuum can be ruled out as we checked a new position on the sample that was neither irradiated by laser or x-ray after the laser exposure measurements, and found that the atomic ratio to be N: I: Pb= 1.03:3.27:1.00, almost the same as the initial ones of laser irradiated and non-irradiated position.

The amount of reduction of N and I by laser radiation along can be calculated from comparing the ratios of irradiated and non-irradiated positions presented in Figure 6. The N ratio is reduced by 0.38 and that of I is 1.02 by 480 min. of laser irradiation after eliminating the contribution from the probing x-ray. Given the stoichiometric ratio of $CH_3NH_3PbI_3$ and that the metallic Pb is now of the total, we can deduce that within the uncertainty of the measurement, about 33 % of the perovskite is decomposed at saturation, and all the volatile components left the surface as described in Eqs. 1 and 2. The total light fluence is $1.96 \times 10^8$ J/m$^2$ at the saturation. The reason of the saturation may be that the exciton quenching centers formed by the metallic lead have effectively quenched all the excitons in the thin film. The formation of the metallic lead requires the sublimation of volatile components, which is kinematically limited by the diffusion process. The decomposition process is then expected to start from the very surface of the perovskite.

If we take a simplified model that the outer layer is completely decomposed while the underlying one remains intact, we can estimate the thickness of the decomposed layer, which gives the lower limit of the decomposed region. From the exponential attenuation, the XPS intensity $I_d$ from a given thickness $d$ is

$$I_d = \int_0^d I_0 e^{-x/\lambda} dx = I_0 \lambda \left(1 - e^{-d/\lambda}\right) \quad (3)$$

where $I_0$ corresponds to the unit emission intensity density and the electron mean free path. The ratio of the intensity from the decomposed layer of thickness $d_1$ to that of the whole layer of thickness $d$ is given by

$$\frac{I_{d1}}{I_d} = \frac{1 - e^{-d_1/\lambda}}{1 - e^{-d/\lambda}} \quad (4)$$

We can deduce that :

$$d_1 = -\lambda \ln\left[1 - \frac{I_{d1}}{I_d}\left(1 - e^{-d/\lambda}\right)\right] \quad (5)$$

With $d$ = 10 nm (the film thickness), = 2.0 nm, and the intensity ratio of metallic Pb and the total Pb, $I_{d1}/I_d$ = 0.33, the degradation thickness $d_1$ caused by laser irradiation is calculated to be 0.82 nm. Our results may also explain the discrepancy between the PL investigation of the perovskite degradation by light. In a layered geometry, the light-induced decomposition is l to the outer surface and the bulk of the material remains intact, thus no spectral shift.[26] On the other hand, nanocrystals of perovskite have much larger relative surface area, and the light-induced degradation is fast and the PL spectral shift continuous as the diffusion process no longer a limiting factor.[27]

A natural question is whether the changes by irradiation reported here are due to heating by the laser beam, as perovskite is known to be unstable above 85 °C even in inert atmosphere.[34] We have carefully analyzed such a possibility and found it unlikely. In our estimate, the gold layer (100 nm) on the silicon substrate acted as a thermal reservoir for the thin perovskite film, and should have prevented any appreciable heating and corresponding heating degradation despite perovskite's extremely low thermal conductance. Following Abbott's model for the heating effects due to a low power CW laser as we used in the experiment, we calculated a steady state temperature increase of less than one degree kelvin.[35] We want to point out that the temperature increase is small because the film is very thin, and appreciable heating may be possible in thicker films or single crystals.[35,36] The calculation demonstrates that the decomposition was caused by the irradiation instead of laser heating effects.

## 4. CONCLUSIONS

We have investigated the effects of laser irradiation on co-evaporated $CH_3NH_3PbI_3$ thin films. After exposing to irradiation for only 120 minutes by a blue laser with wavelength of 408 nm and illumination intensity of $6.8 \times 10^{-3}$ W/mm$^2$, the film began to decompose. The VBM and the core level shift towards higher binding energy, indicating an n-type doping during the irradiation. The appearance of metallic Pb and reduction of N and I demonstrate that the film have decomposed by the laser irradiation. The decomposition saturated after about 480 minutes of laser irradiation saturation, the ratio between metallic Pb and total Pb is about 33%. Correspondingly, N and I are also reduced by the same fraction from the laser irradiation. A model of the decomposition is proposed and a lower limit estimate sets the decomposition at 1 nm of the total 10 nm thick film. SEM shows that the surface was roughened and was no longer continuous after irradiation because of the stress change caused by the degradation. The research indicates that light can cause and accelerate the decomposition of $CH_3NH_3PbI_3$ films, and that keeping the volatile deposition products from leaving the surface is needed for the long term application of perovskite solar cells.

## ACKNOWLEDGMENT

Authors would like to acknowledge the support of the National Science Foundation Grant No. CBET-1437656 and DMR-1303742. Y.L and X.X acknowledge the support of China Scholarship Council. J.H acknowledges the financial support from National Science Foundation under the award of OIA-1538893. Technical supports from the Nanocenter in University of Rochester are highly appreciated.

## AUTHOR INFORMATION

**Corresponding Author**

*E-mail: ygao@pas.rochester.edu

**Notes**

The authors declare no competing financial interest.